\documentclass[aps,pre,superscriptaddress,twocolumn,showpacs,10pt]{revtex4-1}
\def\Title{Quantum and Classical Superballistic Transport in a Relativistic Kicked-Rotor System}
\usepackage{graphicx}
\usepackage{amsmath,amsfonts,amssymb,dsfont,color,bbm,xcolor}
\usepackage{array}
\usepackage{bbold}
\usepackage{natbib}
\usepackage{multirow}

\usepackage[colorlinks]{hyperref}
\newcommand{\avg}[1]{\overline{#1}}
\newcommand{\expct}[1]{\left\langle#1\right\rangle}

\newcommand{\be}{\begin{equation}}
\newcommand{\ee}{\end{equation}}

\newcommand{\rmd}{\mathrm{d}}

\begin{document}

\bibliographystyle{apsrev4-1}

\author{Qifang Zhao}
\affiliation{Department of Physics and Centre for Computational Science and Engineering, National University of Singapore, Singapore 117546, Republic of Singapore}

\author{Cord A.\ M\"uller}
\affiliation{Centre for Quantum Technologies, National University of
Singapore, Singapore 117543, Republic of Singapore}
\affiliation{Department of Physics, University of Konstanz, 78457
  Konstanz, Germany}
\author{Jiangbin Gong} \email{phygj@nus.edu.sg}
\affiliation{Department of Physics and Centre for Computational Science and Engineering, National University of Singapore, Singapore 117546, Republic of Singapore}
\affiliation{
    NUS Graduate School for Integrative Sciences and Engineering,
    Singapore 117456, Republic of Singapore}
\title{\Title}
\date{\today}

\begin{abstract}
As an unusual type of anomalous diffusion behavior, superballistic transport is not well known but has been experimentally simulated recently. Quantum superballistic transport models to date  are mainly based on connected sublattices which are constructed to have different properties.  In this work,
we show that both quantum and classical superballistic transport in the momentum space
can occur in a simple periodically driven Hamiltonian system, namely, a relativistic kicked-rotor system with a nonzero mass term.
The nonzero mass term essentially realizes a junction-like scenario: regimes with low or high momentum values have different dispersion relations and hence different transport properties.  It is further shown that the quantum and classical superballistic transport should occur under much different choices of the system parameters.  The results are of interest to studies of anomalous transport, quantum and classical chaos, and the issue of quantum-classical correspondence.

\end{abstract}

\maketitle
\section{Introduction}
The rich 
transport behavior in complex systems is an important research topic
in statistical physics \cite{review1,review2,review3,review4,PRLbarkai,exponential}. Consider the mean square of a physical quantity (such as position) as a function of time $t$ for an ensemble of particles. For normal diffusion,  this mean quantity is proportional to $t$; whereas for
anomalous diffusion, the mean quantity goes like $\sim t^\nu$ ($\nu\ne 1$), with subdiffusion referring to cases of $0<\nu<1$ and superdiffusion referring to cases of $1<\nu<2$.
In the classical domain, known examples of anomalous diffusion include Brownian motion and heat conduction.
In the quantum domain, wavepacket spreading in a periodic potential leads to ballistic
transport in the mean square position ($\nu=2$).  By contrast, wavepacket spreading in
a quasi-periodic potential often induces subdiffusion or superdiffusion \cite{abe,piechon,ketzmerick}.

The special class of diffusion with $\nu>2$, which may be termed as
``superballistic transport'', is however not as well-studied as 
other cases of anomalous transport behavior.  In the classical domain,
superballistic transport was observed for Brownian
particles~\cite{Bao2007,Siegle2010}.
In the quantum case, a time-dependent random potential was demonstrated to
cause superballistic transport using paraxial optical setting~\cite{Levi2012}.
More related to this work, earlier superballistic transport was found in the dynamics of
wavepacket spreading in a tight-binding lattice
junction~\cite{Hufnagel2001,Zhenjun2012}.   Remarkably,  such type of quantum
superballistic transport was recently experimentally realized by use
of optical wave packets in a designed hybrid photonic lattice
setup~\cite{Stutzer2013}.

The main objective of this work is to use a relatively simple model
system to better understand the difference and connection between
quantum and classical superballistic transport in purely Hamiltonian systems.
To our knowledge, the
model studied in this work represents 
the only dynamical model that can possess superballistic transport in both
quantum and classical Hamiltonian dynamics.  This will shed more
light 
on various mechanisms of superballistic transport as well as on
the general issue of quantum dynamics in classically chaotic
systems. Further studies regarding the subtle correspondence between
quantum and classical superballistic transport can be also motivated.

Specifically,  we consider a relativistic variant
~\cite{Chernikov1989,Matrasulov2005} of the well-known kicked-rotor
(KR) model~\cite{CasatiBook1995} and reveal the quantum and classical
superballistic transport dynamics in the momentum space.  Such a
system can be also regarded as a periodically driven Dirac system, and
it should be of some experimental interest due to recent advances in
the quantum simulation of Dirac-like particles. In the massless case
in which the kinetic energy of a relativistic particle is a linear
function of momentum, the relativistic KR variant was known as the
``Maryland model'', first investigated
by Grempel {\it et. al}~\cite{Grempel1982,Prange1984},
Berry~\cite{Berry1984} and Simon~\cite{Simon1985} to analytically
understand the issue of Anderson localization.  For a Dirac particle
with a nonzero mass, the bare dispersion relation now lies between
linear and quadratic:  for low momentum values the dispersion is
almost quadratic and for very high momentum values the dispersion
approaches a linear function.  In effect, this realizes a situation,
now in the momentum space, in which two (momentum) sublattices with
different dispersion relations (and hence different nature of on-site potential) are connected as a
junction~\cite{Hufnagel2001}.
As shown later, this indeed induces superballistic transport in both
the classical and quantum dynamics.  Interestingly, the detailed
mechanism in the former is still markedly different from that in the
latter. In particular, in the classical case, it is necessary to break
the global KAM curves in the classical phase space because the
superballistic transport roots in an unusually complicated escape from
a phase space regime of random motion to a simple ballistic
structure. By contrast, in the quantum case, breaking global KAM
curves are not essential for quantum superballistic transport to occur
thanks to quantum tunneling through the KAM curves.

This paper is organized as follows. In Sec.~II, we
introduce our model and discuss its relation with the well-known
Maryland model. In Sec.~III we study quantum superballistic dynamics,
followed by a parallel study of classical superballistic transport and
the associated classical phase space structure in Sec.~IV.
Section V concludes this work.

\section{Relativistic Kicked Rotor as a Driven Dirac System}
\label{sec:KRR}
Consider a one-dimensional relativistic quantum KR \cite{Matrasulov2005}:
\begin{equation}
   \label{KickedR}
   H=2\pi\alpha \hat{p} \sigma_x  + M\sigma_z + K\cos (q{\theta}) \sum_{n=-\infty}^{+\infty}{\delta (t-n)},
\end{equation}
where all the variables are scaled and hence in dimensionless units.
Here $\sigma_x$ and $\sigma_z$ are Pauli matrices, 
$v=2\pi \alpha$ represents the speed of light, $M$ represents the static mass energy, $K$ represents the strength of a delta-kicking field that is $(2\pi/q)$-periodic in the coordinate $\theta$ ($q$ is an integer) and unity-periodic in time,  $\hat{p}=-i \hbar_{\text{eff}} \frac{\partial}{\partial \theta}$, where $\hbar_{\text{eff}}$ is a dimensionless effective Planck constant.

In the case of a vanishing $M$, the Hamiltonian~\eqref{KickedR} can be decoupled
into two independent Hamiltonians, each associated with one eigen-spinor of $\sigma_x$. They
are nothing but the so-called Maryland model ~\cite{Grempel1982}:
\begin{equation}
   \label{Maryland}
   H_M= \pm 2\pi\alpha \hat{p} + K\cos (q {\theta}) \sum_{n=-\infty}^{+\infty}{\delta (t-n)}.
\end{equation}
Early studies on this massless relativistic KR~\cite{Grempel1982,Berry1984} investigated the consequences
of rational or irrational values of $\alpha$. Indeed, by mapping the Maryland model onto
a one-dimensional Anderson model, it becomes clear that $H_M$ with an irrational $\alpha$ should display Anderson localization in the momentum space~\cite{Grempel1982}, whereas $H_M$ with a rational
$\alpha=r/s$ ($r$ and $s$ integers) should show ballistic transport, so long as the parameter $q$ in the kicking potential
is an integer multiple of $s$~\cite{Berry1984}. This condition is called a resonance condition.
When this resonance condition is not fulfilled, i.e., $q\neq ns$, it can be
shown that the time-evolving wavefunction of the system (with rational $\alpha$)
repeats itself after every $s$ kicking periods,
so neither Anderson localization nor ballistic transport occurs.
Roughly speaking,
what is known in the standard quantum KR \cite{CasatiBook1995} applies
also here, concerning the importance of the arithmetic nature of $\alpha$ for the dynamics of the Maryland model,
as well as the ballistic transport due to quantum resonances therein.

Though in this work we will focus on cases with a nonzero $M$,
the above-mentioned results for the Maryland model do guide us when it comes to
choose interesting parameter regimes. For example, we shall pay special attention to
whether or not $\alpha$ is a rational value, and
if the kicking potential is 
resonant or not. 
As it turns out, the quantum dynamics is
most interesting in cases with rational $\alpha$ and under the resonance condition.

Since our model is a periodically driven system, we write down its Floquet
operator in the $\theta$-representation, i.e., the propagator associated with one-period time evolution:
\begin{equation}
      \label{FloquetR}
			{U} = e^{-i \frac{1}{\hbar_{\text{eff}}}(2 \pi \alpha \sigma_x \hat{p} + \sigma_z M)} e^{-i \frac{K}{\hbar_{\text{eff}}} \cos (q{\theta})}.
\end{equation}
Without loss of generality we set $\hbar_{\text{eff}}=1$ throughout, keeping in mind that if
$ \hbar_{\text{eff}}\ne 1$, we may just absorb it into other parameters $M$ and $K$. Due to this choice,
with periodic boundary conditions in $\theta$, momentum can only take integer values.
The above expression of the Floquet operator is a product of two exponentials.
The second factor $\text{exp}[-i K \cos (q{\theta} )]$ comes from the kicking field, giving rise to
the hopping between different momentum states (different sites in the momentum space),
whereas the first factor $\text{exp}\left[-i (2 \pi \alpha \sigma_x \hat{p} + \sigma_z M)\right]$
is responsible for generating momentum-dependent on-site phases. In particular,
the state at a site $p$ can be decomposed into local
spin-up and spin-down components, each component acquiring a phase factor $\Phi_p$:
\begin{equation}
   \label{Phase}
   \Phi_p = \text{exp}\left[\mp i \sqrt{(2\pi\alpha p)^2 + M^2}\right].
\end{equation}

Clearly then, it is the periodic and alternative on-site phase accumulation $\Phi_p$ and the hopping
$\text{exp}\left[-i K \cos (q{\theta})\right]$ that determine the quantum dynamics.
This motivates us to also consider a slightly different Floquet
operator ${U}_2$: 
\begin{equation}
      \label{FloquetRN}
			{U}_2 = e^{-i \frac{1}{\hbar_{\text{eff}}} \sqrt{(2 \pi \alpha \hat{p})^2 + M^2} } e^{-i \frac{K}{\hbar_{\text{eff}}} \cos (q{\theta})}.
\end{equation}
For this spinless Floquet operator, the 
two spin components are decoupled.
 Nevertheless, it still contains the same local phase accumulation given by
 $\Phi_p$ and the same hopping term as in our original model described by ${U}$. As seen later,
 it does possess
the essential properties of ${U}$ and as such our physical analysis can be reduced.  More importantly,   the system ${U}_2$ does not have the spin degree of freedom, so its classical limit can be constructed with ease, with the classical Hamiltonian given by
\begin{equation}
   \label{SpinlessH}
   H_C= \sqrt{(2\pi\alpha {p})^2 + M^2} + K\cos (q{\theta}) \sum_{n=-\infty}^{\infty}{\delta(t-n)}.
\end{equation}
Indeed, the so-called classical relativistic KR map studied in the literature ~\cite{Chernikov1989,Nomura1992} was based on such a spinless classical Hamiltonian.
In the following, we study the quantum dynamics using both ${U}$ and ${U}_2$, and the classical dynamics based on $H_C$.



\section{Quantum dynamics}
\label{sec:TH}
The dynamics of a quantum relativistic KR  was previously studied in Ref.~\cite{Matrasulov2005} for relatively short time scales. By extending to a longer time scale and choosing the right parameter regime, quantum superballistic transport is found for driven systems described by ${U}$ as well as its spinless version ${U}_2$. To justify our choices of the system parameters
we first examine cases with an irrational value of $\alpha$.

\subsection{Dynamical localization for irrational \texorpdfstring{$\alpha$}{TEXT} }
\label{sec:IR}
In the 
Maryland model, 
an irrational $\alpha$ leads to localization in the momentum space.
So it is interesting to first investigate how a nonzero mass $M$ changes this
picture.
When $\alpha$ is irrational, the previously defined phase factor $\Phi_p$ in Eq.~\eqref{Phase}
is in general a pseudo-random function of the momentum site. This is different from the Maryland model,
in which $M=0$ and the corresponding $\Phi_p$ would then reduce to a quasi-periodic function of momentum sites. In the light of the mapping from a KR system to the Anderson localization model \cite{Fishman}, this seems to indicate that a non-zero $M$ favors dynamical localization in the momentum space. Note also that,
for very large values of $p$, the relative importance of the $M$ term in $\Phi_p$ will diminish, and then effectively a Maryland model will be recovered and dynamical localization is still guaranteed~\cite{Grempel1982}.  Thus, in the entire momentum space,  a non-zero $M$
is expected to strengthen the dynamical localization, thus also wiping out any possibility of anomalous diffusion or superballistic transport.

\begin{figure}
\centering
   \includegraphics[width=\linewidth,clip=true]{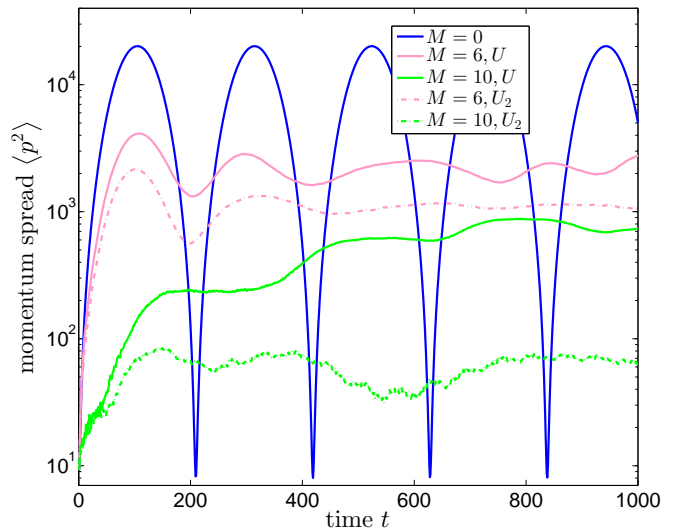}
   \caption{(Color online) Momentum spread $\expct{p^2}$ as a function
     of number of kicks (time $t$).
	  Solid lines represent results generated by the Floquet
          operator ${U}$ defined in Eq.~(\ref{FloquetR}), with the
          initial state given by spin-up
and a Gaussian wavepacket $\sim \exp (-p^2/2\sigma_p^2)$.  The dot-dashed lines represents the results generated by a spinless Floquet
		operator ${U}_2$ defined in Eq.~(\ref{FloquetRN}),
                with the initial wavepacket 
given 
by the  Gaussian
$\sim \exp (-p^2/2\sigma_p^2)$. In both
		cases, $\sigma_p=4$, irrational $\alpha=1/3+0.01/2\pi$, $K=0.8$, $q=3$ and $M$ takes values of
 0, 6 and 10. (Results of ${U}$ and ${U}_2$ with $M=0$ are identical.) $\expct{p^2}$ is seen to be strongly localized in all the examined examples. Here and in all other figures, all plotted quantities are in dimensionless units.}
   \label{fig:M_Irra}
\end{figure}

Results of our numerical simulations presented in Fig.~\ref{fig:M_Irra} support our above view.
For both the full Floquet operator ${U}$ and its spinless version ${U}_2$, it is seen
from Fig.~\ref{fig:M_Irra} that the momentum spread $\expct{p^2}$ decreases
as $M$ increases, i.e., a larger $M$ enhances dynamical localization.

With the same set of system parameters,  the mean momentum spread
under the evolution of the Floquet operator ${U}$ is similar to that under the evolution of its spinless
version ${U}_2$.  This confirms that the common on-site phase accumulation
function $\Phi_p$ has captured the main features of dynamical localization.
It is also interesting to comment on the differences between these two cases. That is,
$\expct{p^2}$ associated with ${U}$  is always larger than that associated with ${U}_2$.
As such, the spin degree of freedom is seen to slightly weaken dynamical localization.
This is somewhat expected.  Indeed, the spin degree of freedom introduces
two channels for the dynamics and a multi-channel Anderson model does increase
the localization length~\cite{Mello1988}. We note in passing that the spin degree of freedom can even
cause Anderson transition in two-dimensional disordered systems~\cite{Evangelou1987,Sheng1996}.
Certainly, as $M$ increases, this spin effect should decrease because it becomes more costly in energy for the two spin channels to interact.

Finally, we mention the benchmark result in Fig.~\ref{fig:M_Irra} for the $M=0$ case that represents the Maryland model. The perfect revival of $\expct{p^2}$ was predicted by Berry~\cite{Berry1984}.  Using our system parameters depicted in the caption of Fig.~\ref{fig:M_Irra},
Berry's result gives $\expct{p^2}(t) \sim \sin^2 0.015t /\sin^2 0.015$ and our simulation agrees with this.

\subsection{Superballistic transport for rational \texorpdfstring{$\alpha$}{TEXT} and on-resonance potential}
As we already discussed in the previous subsection, for large momentum values,
the effect of a nonzero mass term $M$ will diminish and effectively the Maryland model will re-emerge.
So if we were not to choose an on-resonance potential, then in regimes of large momentum
quantum revivals should occur for 
rational $\alpha$, which is not of interest here.  In addition, we
also observed that for an off-resonance kicking potential and for 
rational $\alpha$, a non-zero $M$ further suppresses the already bounded momentum spread.
With these understandings, it is clear that we should step into the interesting situation where $\alpha=r/s$ is rational and
the kicking potential $K\cos (q \theta)$ is on resonance, i.e, $q=ns$. For convenience we
choose $q=s$.  Reference~\cite{Matrasulov2005} computationally
investigated exactly the same situation, but it was argued therein that the phase factor
$\Phi_p$ as a pseudo-random function of momentum should suffice to
localize the momentum spread. As we show below, both qualitatively and
quantitatively, 
this claim is correct only for 
low momentum values, and overall a much richer transport behavior can be found.


Let us start with the Maryland model for which $M=0$. Then the
momentum space is translational invariant with 
period $s$. As such, states will in general spread ballistically (the off-resonance case is an exception). In our case, $M\neq 0$ and within each period, the on-site phase $\Phi_p$
acquired by the system becomes a quasi-random function of $p$, thus
dynamical Anderson localization or suppression of momentum spread is
expected.  However, as momentum 
increases, the nonzero $M^2$ term in $\Phi_p = \exp\left(\pm i[\sqrt{(2\pi\alpha p)^2+M^2}]\right)$ becomes less important as compared with $(2\pi\alpha p)^2$.  For very large momentum, the $M^2$ term represents a very weak perturbation and hence the dynamics should resemble that of the Maryland model.

The above qualitative analysis makes it clear that the overall
dynamics depends on many factors. On a sublattice representing low
momentum values from $-p_c$ to $+p_c$, dynamical localization takes
place and the system is effectively in a disordered regime.  On a
sublattice representing higher momentum values, ballistic transport is
expected and the system is effectively in a periodic regime.  Whether
or not a state is localized or delocalized now depends on where it is
initially located, and on the size of the disordered regime as
compared with the localization length. For example, if an initial
state is localized at the center of the disordered regime and if the
localization length is much shorter than 
the disordered
region, then the system may be trapped there for an extremely long
time.  On the other hand, if the initial state is already located
close to the high-momentum sublattice (closeness is with respect to
the localization length), then as the kicking field induces population
transfer between the disordered sublattice and the periodic
sublattice, the system will be quickly delocalized.
In this sense, our model, through its natural dispersion relation,
realizes a lattice junction analogous to that considered in
Refs.~\cite{Hufnagel2001,Zhenjun2012}.   Certainly, in our model here
there is no sharp transition between the two qualitatively different
sublattices, but the critical momentum value is expected to scale with
$M/(2\pi \alpha)$.

\begin{figure}
\centering
   \includegraphics[width=\linewidth,clip=true]{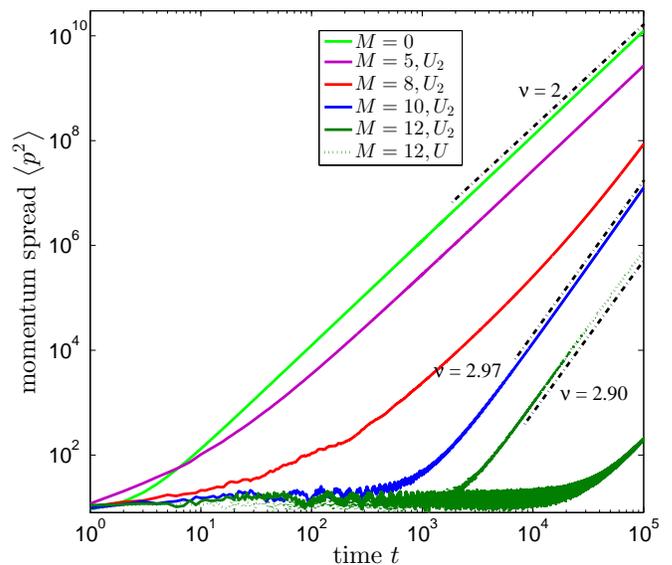}
   \caption{(Color online) Ensemble averaged momentum spread $\expct{p^2}$ vs time, for $M=0$, $M=5$, $M=8$, $M=10$ and $M=12$ (from top to bottom). Results with $M=5$, $M=8$, $M=10$ are for the spinless Floquet operator ${U}_2$ defined in Eq.~(\ref{FloquetRN}). For $M=12$, results for both ${U}$ (the one fitted by a power-law with $\nu=2.90$)
   and ${U}_2$ (the most localized case) are plotted. (Result with $M=0$ is applicable to both ${U}$ and ${U}_2$.) Other system parameters are the same as in Fig.~\ref{fig:M_Irra}, except for $\alpha=1/3$.  The two dashed straight lines represent a power-law fitting $\expct{p^2} \propto t^{\nu}$ (for a certain time window) with $\nu=2.90$ or $\nu=2.97$, indicating quantum superballistic transport with an exponent close to $\nu=3$.}
   \label{fig:M_Ra_Log}
\end{figure}

Representative results from our numerical experiments are presented in
Fig.~\ref{fig:M_Ra_Log}.  There it is seen that as $M$ increases, an initial state localized
at the center of the disordered sublattice will be trapped for a longer period. This is consistent with
our understanding that an increasing $M$ leads to a longer disordered sublattice as
well as a shorter localization length.
 Other numerical results (not shown)
show that the values of
$\alpha$, the kicking strength $K$, and the potential parameter $q$ can all affect the duration
during which an initial state is trapped in the disordered regime.

Two of the computational examples shown in Fig.~\ref{fig:M_Ra_Log} also display quantum superballistic transport,
i.e. $\expct{p^2} \propto t^{\nu}$ with $\nu>2$, where $t$ is the number of kicks.
This can now be explained using the idea from Ref.~\cite{Hufnagel2001}.
In particular, assuming that the length of the disordered momentum sublattice is
larger than the corresponding localization length, then the disordered regime
serves as a source to provide slow probability
leakage into a periodic regime at an almost constant rate. Then we have
\begin{equation}
     \label{psquare}
		 \expct{p^2}(t) \approx B + a C t^2 + a \int^t_0 { R(t') (t-t')^2 \rmd t'}.
\end{equation}
Here $R(t)$ is the probability leaking rate from the disordered sublattice to the periodic sublattice,
$B$ represents the contribution from the disordered regime, which is almost constant and can be neglected.
$C$ represents the probability of the initial state already placed in the periodic regime, which is
0 due to our choice of the initial state located at the center of the disordered regime. $a$ characterizes the ballistic transport coefficient, which depends on many system parameters.
If we approximate $R(t)$ by a constant $\Gamma$, then  from Eq.~\eqref{psquare} we approximately have
\begin{equation}
     \label{psquare2_Q}
		 \expct{p^2}(t) \propto  a \Gamma  t^3.
\end{equation}

 As reflected by the two cases shown in Fig.~\ref{fig:M_Ra_Log}, namely, the case
 of $M=12$ for the full Floquet operator ${U}$ and the case of $M=10$ for the spinless Floquet operator ${U}_2$ defined in Eq.~(\ref{FloquetRN}),  $\expct{p^2} \propto t^{\nu}$, with $\nu\approx 2.90$ or $\nu\approx 2.97$, for a very long time scale and covering a huge range of $\expct{p^2}$ (note the logarithmic scales used in the plot).  These two superballistic exponents
are very close to $\nu=3$, in agreement with the above theory.
For the same two cases,  we have set $p_c \approx 100 M/(2\pi\alpha)$ and
record the probabilities inside $[-p_c, p_c]$ as a function of
time. This probability indeed decreases linearly with time.  This
further confirms the physical mechanism behind the quantum
superballistic transport seen here.  The case shown in
Fig.~\ref{fig:M_Ra_Log} with $M=5$ displays ballistic transport as the
case of $M=0$. This is so 
because for a small value of $M$ the
initial state quickly experiences ballistic transport on the clean
sublattice.  For the intermediate 
case $M=8$, the momentum spread does not
show any clear power-law dependence. In this transitional case, the
localization length and the size of disordered lattice are  
comparable and hence the leakage from the disordered sublattice to the periodic
sublattice occurs 
no longer at an almost constant rate.  We stress that
the shown cases represent but a few examples. 
Many similar results of
quantum superballistic transport are obtained for both the full
Floquet operator ${U}$ involving two spin channels and for the spinless
Floquet operator ${U}_2$.

Though the quantum superballistic transport here is explained in the
same manner as in Ref.~~\cite{Hufnagel2001}, we stress that the
effective two-sublattice configuration is not artificially designed. Rather, it
emerges as a natural consequence of the relativistic dispersion relation with a
nonzero mass.



\section{Classical dynamics}
\subsection{Classical phase space structure}
In this subsection we will study the classical relativistic KR described by
the Hamiltonian in Eq.~\eqref{SpinlessH}. To that end, it is necessary
to examine the
phase space structure, which can be generated from the relativistic standard map~\cite{Chernikov1989,Nomura1992}.  In particular,
the states right after the $N$-th and $N+1$-st
kick are connected by the map
\begin{equation}
\label{RSM2}
\begin{split}
   \theta_{N+1} &= \frac{v^2 p_N}{\sqrt{v^2 p_N^2 +M^2}} + \theta_N\ \  (\text{mod } 2\pi/q), \\
   p_{N+1} &= q K \sin (q \theta_{N+1}) + p_N.
\end{split}
\end{equation}



\begin{figure}
\centering
   \includegraphics[width=\linewidth,clip=true]{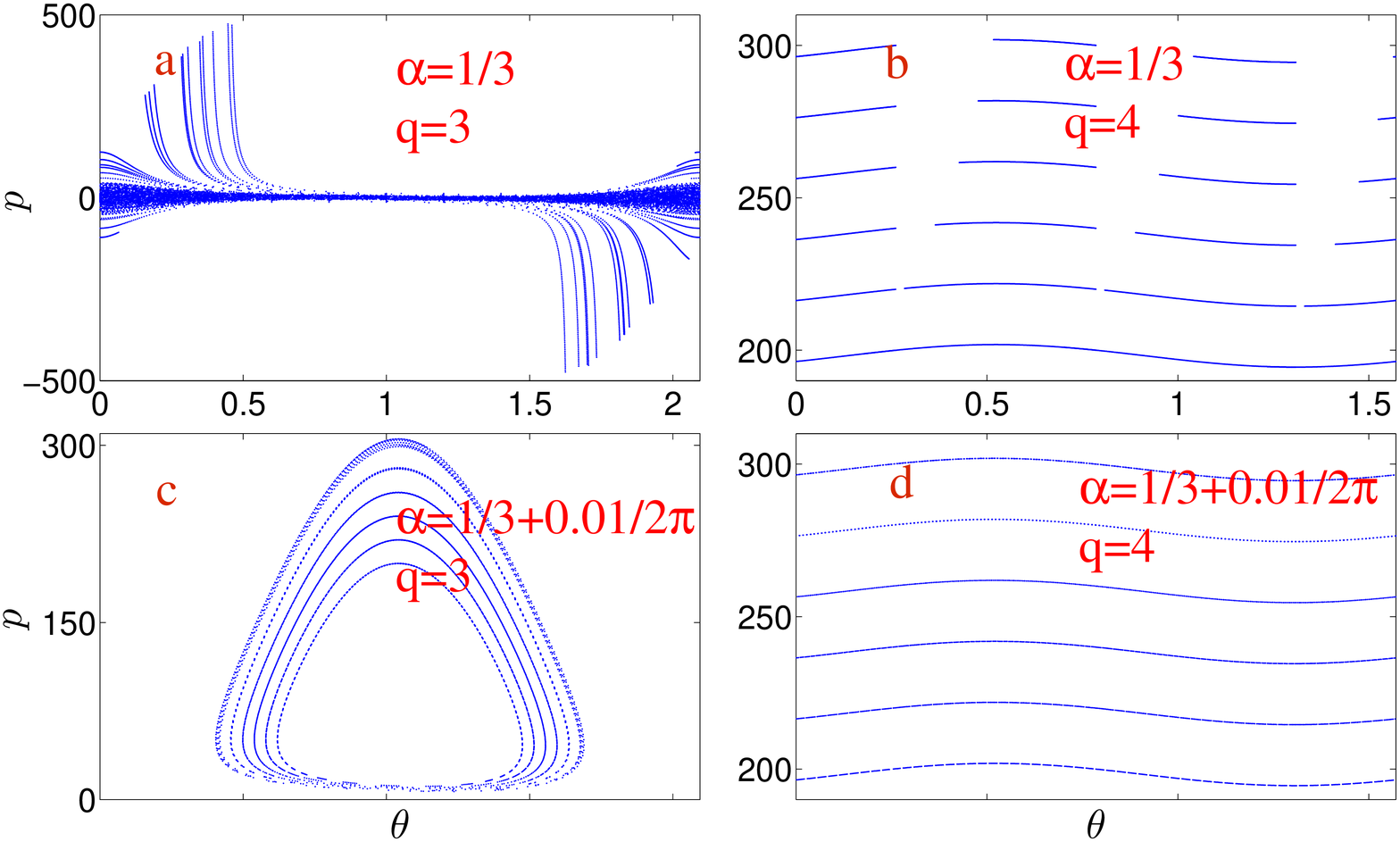}
   \caption{(Color online) Phase space structure of $H_C$ defined in Eq.~(\ref{SpinlessH}), for four different situations, namely, (a) rational $\alpha$ and on-resonance kicking potential, (b) irrational $\alpha$ with off-resonance kicking potential, (c) irrational $\alpha$ with
   an almost on-resonance kicking potential; and (d) irrational
   $\alpha$ with an off-resonance kicking potential. $M=10$ and
   $K=1.6$.  The phase space invariant curves are seen to be unbounded
   in momentum in panel (a), but bounded
    in panels (b)-(d).}

   \label{fig:PhaseSpaceAllCases}
\end{figure}

In the quantum case, either an irrational $\alpha$
or a non-resonant kicking potential causes localization.  Considering quantum-classical correspondence,
this suggests bounded (i.e., localized in momentum) invariant curves in the phase space.
Results in Fig.~\ref{fig:PhaseSpaceAllCases}(b)-(d) support this view.

To understand the phase space structure,  we first recall the Maryland model~\cite{Berry1984}, which can approximately describe the dynamics for sufficiently large
momentum values. That is, if $p$ is large, then we again neglect the $M$ term in the Hamiltonian. Then the mapping
in Eq.~\eqref{RSM2} (after dropping the $M$ term) reduces to the mapping associated with the Maryland
model.  For the Maryland model, the following equations hold for either an irrational $\alpha$
or a non-resonant kicking potential:
\begin{equation}
\label{P_theta_bounded}
\begin{split}
   \theta_{N} &= Nv + \theta_0, \\
   p_{N} &= \frac{1}{2}\csc\frac{qv}{2}\left(\cos(\frac{qv}{2}+q\theta_0) - \cos\left[(N+\frac{1}{2})qv+q\theta_0\right]\right)\\
	         &\quad + p_0,
\end{split}
\end{equation}
where as introduced before, $v=2\pi\alpha$.
Clearly then, for an irrational $\alpha$,  the term  $Nv$ 
can densely cover the entire $\theta$ domain $[0,2\pi]$. As such, as $N$ increases, the values of $(\theta_N,p_N)$ fill a complete
sine curve in the phase space.  On the other hand, if $\alpha$ is rational, $q\alpha$ becomes
a fraction under the assumed off-resonance condition, then $\theta_N$ can only take discrete points
in $[0,2\pi]$ such that $p_N$ also takes a few isolated values~\cite{Berry1984}.

These observations for the Maryland model can be used to directly explain the panel (d) in Fig.~\ref{fig:PhaseSpaceAllCases}.  For panel (b) where $\alpha$ is rational but the kicking potential is off resonance, a nonzero $M$ also causes the dynamics to densely fill the entire $\theta$ domain, which constitutes an interesting difference from the Maryland model. However, as expected, after the same number of iterations, regimes with low momentum values can generate a complete phase invariant curve faster, and regimes with high momentum values may still have holes to be filled in.   As to panel (c) where the product of $\alpha q$ is close to an integer, the oscillation amplitude in momentum get larger as the factor $\csc\frac{qv}{2}$ in the map in Eq.~(\ref{P_theta_bounded}) can be a large number.  Putting all these cases together, it is seen that in the three situations represented by panels (b)-(d) of Fig.~\ref{fig:PhaseSpaceAllCases}, KAM invariant curves localized in momentum are the main characteristic of the classical phase space.

\begin{figure}
\centering
   \includegraphics[width=\linewidth,clip=true]{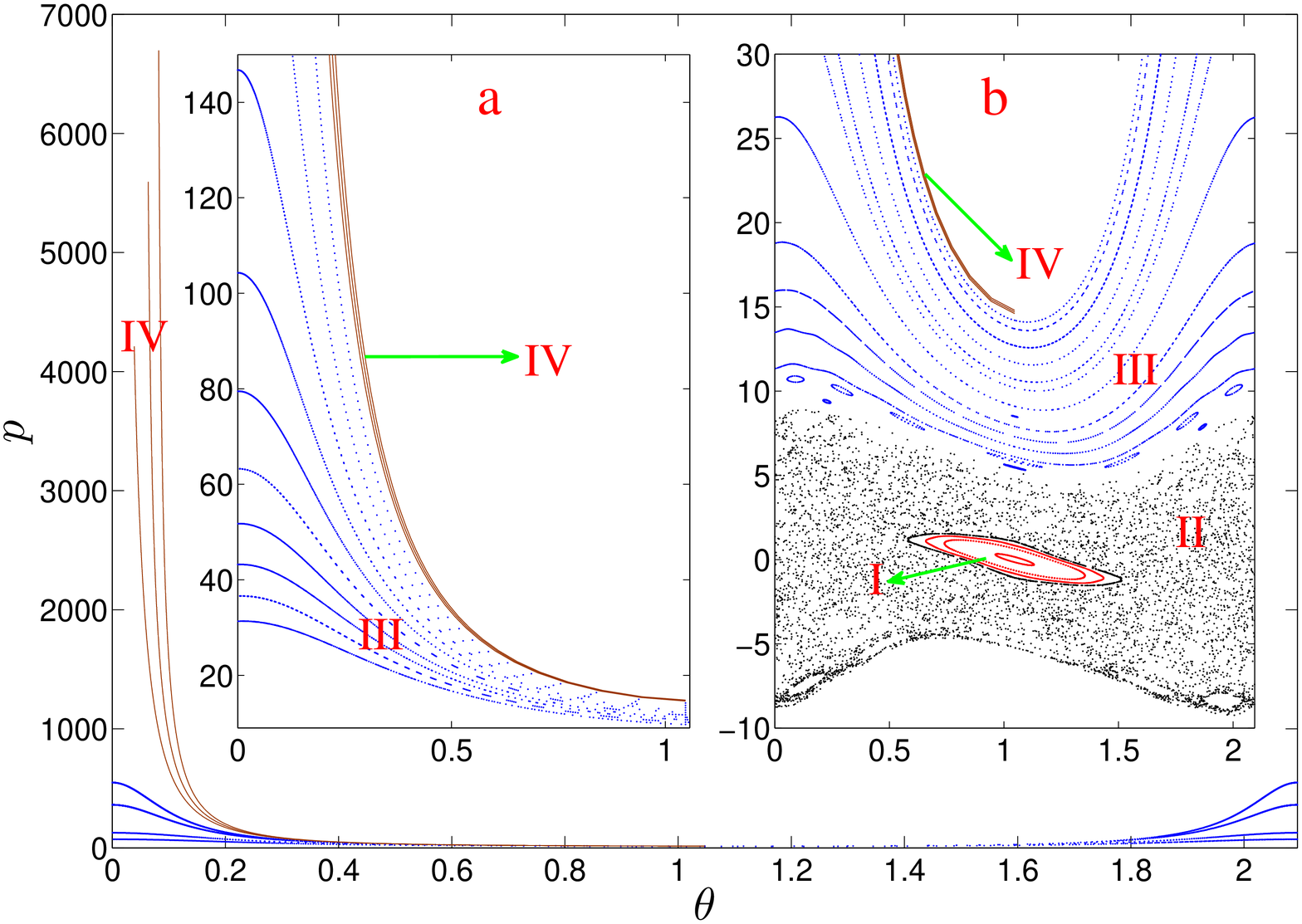}
   \caption{(Color online) Detailed phase space structure for a situation similar to panel (a) in Fig.~\ref{fig:PhaseSpaceAllCases}. Here system parameters are given by
	   $K=0.8$, $M=10$, $\alpha=1/3$, and $q=3$. Panel (a) depicts regimes with large
		momentum values, and panel (b)  depicts regimes with
		low momentum values. Curves in regime I are local KAM curves, KAM curves in regime III
		are global curves and hence localize momentum.
     The black dots regime in II are generated by a single
		initial condition  and indicate a chaotic sea.  Curves in regimes IV that are almost parallel to
the momentum axis are ballistic structures.}
   \label{fig:KAM_detail}
\end{figure}

So now we are left with the last situation in which $\alpha$ is rational and the kicking potential is on resonance. The associated phase space structure is presented in panel (a) of Fig.~\ref{fig:PhaseSpaceAllCases} and in
Fig.~\ref{fig:KAM_detail}. The phase space structure is remarkably complicated and interesting.
To investigate this in detail,
we divide the phase space into four regimes, namely regimes I, II, III and IV
for increasing absolute values of momentum.

Let us take one example to look into the special phase space structure. As seen in Fig.~\ref{fig:KAM_detail} for $K=0.8$, $M=10$, the four regimes have qualitatively different behavior.
In regime I where $p$ is small, local KAM curves dominate.  As $p$ increases, the feature of the phase space become
chaotic in regime II. This is followed by global KAM curves in regime III. These global KAM curves are localized in momentum and  bound the chaotic sea seen in regime II.
Finally, in regime IV for quite large momentum, ballistic curves, which become more and more parallel to the momentum axis, are seen.  These curves are called ballistic curves because
once a trajectory lands on such a structure its momentum variance will evolve ballistically.


The four regimes identified above may not always appear together. Their presence and borders are
determined by $M$ and $K$. To shed some light on this,  one may consider $H_C$,
in the two opposite limits, i.e.,  ``non-relativistic'' $vp \ll M$ limit 
and ``ultra-relativistic'' limit $vp \gg M$. In the first limit, $H_C$ becomes
\begin{equation}
   H_S =  M + \frac{1}{2} \frac{v^2 p^2}{M} + K\cos q\theta \sum_{n=-\infty}^{\infty}{\delta(t-n)},
   \end{equation}
   and in the second limit, $H_C$ assumes
   \begin{equation}
   H_L = v|p| + K\cos q\theta \sum_{n=-\infty}^{\infty}{\delta(t-n)}.
\end{equation}
Regimes I and II can be understood via $H_S$, while regime IV can be well understood by
$H_L$.  In particular, $H_S$ is the conventional kicked rotor, which makes the regular-to-chaotic
 transition as $K$ increases \cite{CasatiBook1995}. $H_L$ is much similar to the Hamiltonian of the Maryland model, which is known to produce ballistic trajectories in the momentum space \cite{Berry1984}.
Indeed, it can be shown that the asymptotic (in the large $p$ limit)
form of the ballistic trajectories are described by
$\theta=\text{constant}$, and they are hence completely 
parallel to the momentum axis.

\begin{figure}
\centering
   \includegraphics[width=\linewidth,clip=true]{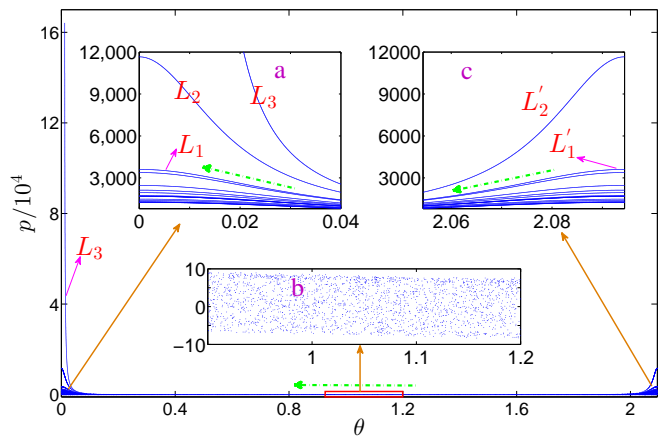}
   \caption{(Color online) Evolution of a single trajectory from $(\theta_0,p_0)=(0,0.7)$.
   The total evolution time is $2\times 10^6$, by which time this trajectory
   becomes clearly ballistic.
    The whole process is remarkably complicated. Panels (a) (b) and (c)
    zoom in some small portions of the shown phase space, with the momentum
    values in the middle-$\theta$ regime seen to be highly localized.
    Green dash-dotted arrow shows the moving direction of the trajectory along
    the ``curves"  traced by the trajectory. These curves shown here emerges
		after $t\approx 7.6\times 10^5$, and similar curves developed in earlier
		time with smaller $p$ values. Note that these curves are not
    KAM invariant curves as the system will eventually leave them.
		We define the motion along these curves as ``transient chaotic motion".
    }
   \label{fig:PhaseSinglePoint}
\end{figure}

As is found from our computational studies, an increase in $K$ may destroy the KAM curves in regimes I and III, and then turns them to a (possibly transient) chaotic sea as well.
An increase in $M$ will generate more KAM curves in the phase space.
The complexity of the phase space perhaps deserves more careful studies. For our purpose here,
we emphasize that for a sufficiently large $K$, the phase space is mainly composed of a seemingly
chaotic sea and ballistic trajectories. It is however
challenging to identify a clear boundary between trajectories eventually landing on the ballistic structure and those always doing random motion.   To appreciate this complexity, in Fig.~\ref{fig:PhaseSinglePoint} we illustrate that once global curves are all broken, how an individual trajectory might eventually land on the ballistic structure after transient, but a long period of, ``chaotic" motion.  The whole process is like the following:  after the system has wandered in the transient chaotic sea [see panel (b)] for a long time, the system finally reaches $L_1$ and keeps moving to the left. Then it reaches $L^{'}_1$ and continues to move towards the left.  It then
passes the central transient ``chaotic sea". Later the system has a chance to arrive $L_2$,
followed by $L^{'}_2$.  The system eventually reach a ballistic curve $L_3$ and then keeps moving up in the momentum space. In brief, it takes 3 stages for a trajectory launched from the regime illustrated in
panel (b) of Fig.~\ref{fig:PhaseSinglePoint} to finally turn to ballistic motion. First, it wanders highly randomly in a transient
chaotic sea. Second, it moves alternatively along some smooth curves [such as those shown in panel (a) and panel (c) in Fig.~\ref{fig:PhaseSinglePoint}], between which the system returns to the transient chaotic sea, but with the overall tendency towards curves of large momentum values.  Lastly, the system evolves on a simple ballistic structure.

\subsection{Classical superballistic transport}

\begin{figure}
\centering
   \includegraphics[width=\linewidth,clip=true]{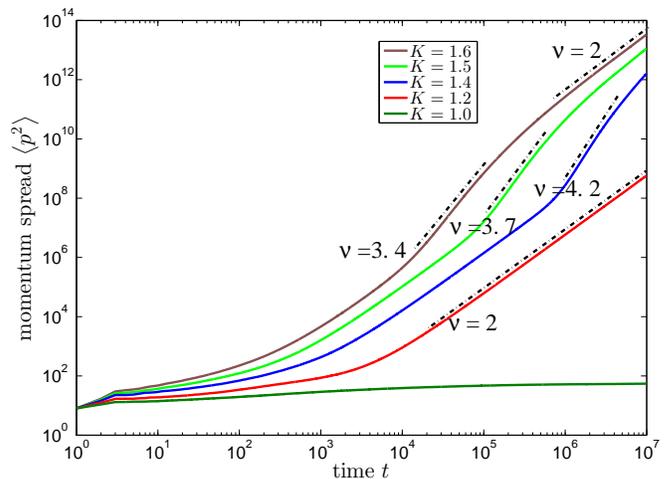}
   \caption{(Color online) Ensemble averaged classical momentum spread $\expct{p^2}$ vs time (the number of kicks).
   The initial conditions are sampled from a Gaussian distribution described by Eq.~\eqref{Distribu_Cla} with $\sigma_p=2\sqrt{2}$
		and $\sigma_{\theta}=1/\sigma_p$, so that the initial phase space distribution is analogous to the quantum initial state used in Fig.~\ref{fig:M_Ra_Log}.
	  Here $v=2\pi/3$, $q=3$, $M=10$, and from top to bottom, $K=1.6, 1.5, 1.4, 1.2$, and $1.0$.
 Note that with other system parameters being the same, quantum superballistic transport
  already occurs for $K=0.8$ (see Fig.~\ref{fig:M_Ra_Log}).  Here the classical superballistic transport emerges until $K$ reaches 1.4.}
   \label{fig:KsM10}
\end{figure}

In our simulations, we always choose $ 10^5$ phase space points randomly sampled from the following Gaussian distribution
\begin{equation}
 \label{Distribu_Cla}
f(p,\theta) = \frac{1}{2\pi \sigma_p \sigma_{\theta}} \text{exp}(-\frac{p^2}{2 \sigma_p^2}) \text{exp}(-\frac{\theta^2}{2 \sigma_{\theta}^2}).
\end{equation}
Note that this Gaussian distribution is analogous to the initial Gaussian wavepacket we used in our
quantum dynamics calculations. We then evolve this ensemble of classical trajectories
according to the relativistic KR map described by Eq.~\eqref{RSM2} and examine
the ensemble averaged $\expct{p^2}$ as a function of $t$, i.e., the number of iterations.
Interestingly, the time dependence of $\expct{p^2}$ is rich, and if we fit $\expct{p^2}$ by the power-law $\sim t^{\nu}$ for appropriate time windows, the exponent $\nu$ can be larger than 2.  In fact, sometimes $\nu$ can be even larger than three or even four.  Some examples are shown in  Fig.~\ref{fig:KsM10}. In one case, the superballistic transport exponent is found to be as large as $\nu=4.2$, for a time window from
$t=10^6$ to $t=3\times 10^6$.
 On the one hand this confirms that the classical dynamics may display superballistic transport, on the other hand it is necessary to better understand the underlying mechanism.  By exploring many parameter choices, it is found that breaking the global KAM curves with an increasing ratio
 $K/M$ is a necessary condition.   This is already a clear difference from quantum superballistic transport.
 For example, for the results shown in Fig.~2, $v=2\pi/3$, $q=3$, $M=10$,  quantum superballistic transport occurs already for $K=0.8$, but in the classical case shown in Fig.~\ref{fig:KsM10}, superballistic transport is observed only when $K$ exceeds 1.4.  Thus, in the quantum case, a global KAM invariant curve does not forbid the population leakage from the classically
 chaotic regime to the ballistic regime, an indication of quantum tunneling.

To further understand the numerical results,
we find it necessary to also account for 
normal diffusion as the trajectories seek to land on ballistic trajectories
from the chaotic sea.  The associated normal diffusion rate is assumed to be $D_0$.   We further assume that the leakage rate from the chaotic sea to the ballistic structure
is given by $R(t)$. Then analogous to Eq.~(\ref{psquare}),  we
expect to have
\begin{equation}
     \label{psquare2_C}
		 \expct{p^2}(t) \approx B + D t + a C  t^2 + a \int_0^t R(t') (t-t')^2 \rmd t',
\end{equation}
where $B=\rho_0 \avg{p^2_0}$, with $\rho_0$ being the
 fraction of trajectories confined in some local stable islands in Regime I
 and $\avg{p^2_0}$ being their average momentum spread;
$D=\rho_1 D_0$, with $\rho_1$ representing the fraction of trajectories undergoing
normal diffusion, with $D_0$ being the associated
diffusion constant; $C=\rho_2$ is the faction of trajectories initially placed on ballistic structures, with $a$ being the diffusion coefficient.  Note that, unlike in the quantum case, at this point
 we do not first assume a constant probability leakage
rate because, as seen below, the leakage involves different behavior at different time windows, and so $R(t)$ can be rather complicated. Indeed, as seen from Fig.~\ref{fig:PhaseSinglePoint}, once the global KAM curves are destroyed, the escape  from a (transient) chaotic sea to a ballistic structure is extremely complicated: the boundary between them
is hard to identify and different initial conditions sampled from an initial Gaussian ensemble may need drastically different times to reach a ballistic structure.

\begin{figure}
\centering
   \includegraphics[width=\linewidth,clip=true]{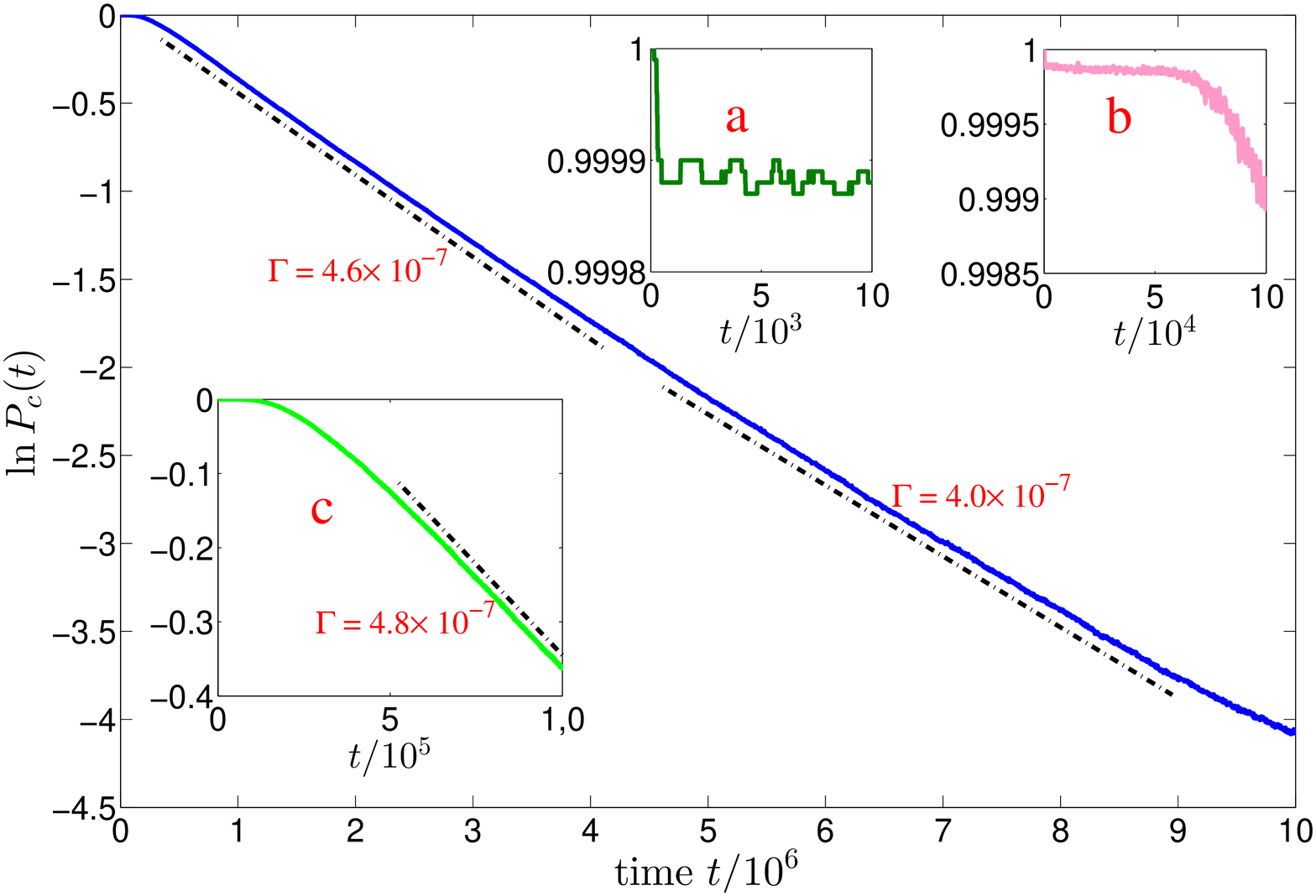}
   \caption{(Color online) Results of $\ln P_c(t)$ vs time,  where $P_c$ defined in the text
    represents the occupation probability in a non-ballistic regime.
    For the two insets (a) and (b), $P_c(t)$ vs time is plotted for two early stages, where
    it can be seen that initially most trajectories are trapped in the initial non-ballistic regime
    for  $t<t_1\approx 5\times 10^4$. Then $P_c(t)$ starts to
		decrease appreciably in a nonlinear fashion. The inset (c) shows that the time dependence
 of $\ln P_c(t)$ is highly nonlinear before $t=t_2$. However, after $t_2\approx 5\times 10^5$,
		 $\ln P_c(t)$ and $t$ display a linear relation.
		The system parameters are the same as the case of $K=1.4$ in Fig.~\ref{fig:KsM10}.}
   \label{fig:KsM10ProbLeak_Combine}
\end{figure}

The coexistence of normal diffusion, ballistic transport, and the potentially complicated leakage rate
$R(t)$ makes the time dependence of $\expct{p^2}(t)$ even more interesting than the quantum case.
Let us roughly define a regime $[-P_0,P_0]$ in the phase space, where
trajectories are not doing ballistic motion. In connection with our
observations made from Fig.~\ref{fig:PhaseSinglePoint}, we choose
$P_0=500$. Let $P_c(t)$ be the occupation probability of this regime
and $P_u(t)=1-P_c(t)$ be the occupation probability on ballistic
structures.  Then $R(t)=-\frac{\rmd P_c(t)}{\rmd t}$.  
$P_c(t)$
is plotted in Fig.~\ref{fig:KsM10ProbLeak_Combine} on either linear or
logarithmic scales,
for different time windows, for the value $K=1.4$ already studied in
Fig.~\ref{fig:KsM10}.
The right inset of Fig.~\ref{fig:KsM10ProbLeak_Combine}
indicates that initially most trajectories are trapped in the regime $[-P_0,P_0]$,
until $t=t_1\approx 5\times 10^4$. Then $P_c(t)$ starts to
decrease more appreciably, with a time dependence not easy to fit [see the first part of the curve in panel (c)].
After $t=t_2\approx 5\times 10^5$ kicks however,
the relation between $P_c(t)$ and $t$ becomes much more evident,
i.e. $\ln P_c(t) \propto -\Gamma t$, which indicates an exponential
decay.  The emergence of an exponential decay suggests 
that the ensemble 
has reached a certain steady configuration, as the escape probability  now becomes proportional to the occupation probability itself.
As also shown by the bottom inset and by the main figure of Fig.~\ref{fig:KsM10ProbLeak_Combine}, the coefficient
$\Gamma$ slightly changes with time.
Expanding such an exponential decay to the first order, this escape would
amount to an almost  constant leakage rate of jumping onto phase space ballistic structures. As a result,
one would naively expect, like our analysis in the quantum part, a superballistic transport case with $\nu=3$. This prediction is certainly oversimplified as compared with our actual results shown in Fig.~\ref{fig:KsM10}.

To better digest the results shown in Fig.~\ref{fig:KsM10},  we again focus
on the case $K=1.4$ in connection with the time dependence of $P_c(t)$ in Fig.~\ref{fig:KsM10ProbLeak_Combine}. In the very beginning, $P_c(t)$ remains almost a constant until $t_1\approx 5\times 10^4$ kicks, so for this time period $R(t)$ is essentially zero.
Therefore initially only the first three terms in Eq.~\eqref{psquare2_C}
are non-zero, suggesting that the diffusion exponent of $\expct{p^2}$ should be less than two. This explains the actual numerical result during the early stage.
The plotted curve in Fig.~\ref{fig:KsM10} at early times also has an increasing slope. This can be
explained as follows.  During the early stage, we have $B>D>aC$.
When $t>B/D$, the $Dt$ term starts to dominate so $\expct{p^2}$ is close to normal diffusion.
Similarly, when $t>D/(aC) \approx 10^3$, the $aCt^2$ term exceeds the first two,
so we have a behavior close to ballistic transport for a quite long period until $t=t_2\approx 5\times 10^5$. On the other hand, from Fig.~\ref{fig:KsM10ProbLeak_Combine}, it is observed that
since as early as $t=t_1\approx 5\times 10^4$, $R(t)$ is already
non-zero. So the leakage to the ballistic regime is building up long before an exponential leakage is observed at $t>t_2$.  This early-stage leakage to the ballistic regime
starts to affect the time dependence of $\expct{p^2}$ only until $t\approx 10^6$. We conjecture that this is the reason why in Fig.~\ref{fig:KsM10} a simple relation $\expct{p^2}(t)\sim t^3$ is not observed.
In addition, the lack of such a simple superballistic behavior with $\nu=3$
is also consistent with the apparent nonlinear time dependence shown in panel (c) of Fig.~\ref{fig:KsM10ProbLeak_Combine} before $t=t_2$.


\begin{figure}
\centering
   \includegraphics[width=1.0\linewidth,clip=true]{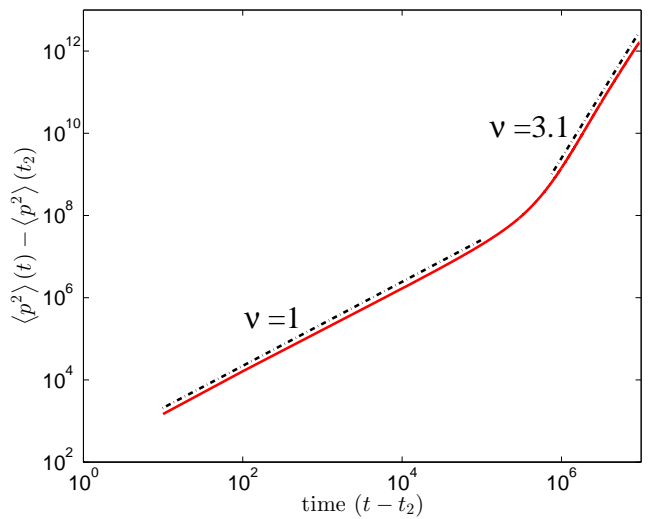}
   \caption{(Color online) $[\expct{p^2}(t) -\expct{p^2}(t_2)]$ vs $(t-t_2)$, an analysis motivated by
    a reset of the start time at $t=t_2$ (see the text for the details).  The result here mainly displays a normal diffusion stage and a superballistic transport stage with an exponent close to $\nu=3$.  The computational example presented here is the same one in Fig.~\ref{fig:KsM10ProbLeak_Combine} with $K=1.4$.  The dashes lines represent power-law fitting.
	  }
   \label{fig:K1_4M10_truncated}
\end{figure}

To confirm our qualitative analysis above, we now redefine the start time as the point
when an exponential decay of $P_c(t)$ can be clearly identified.
Again using the computational example shown in Fig.~\ref{fig:KsM10ProbLeak_Combine},
we now use $t_2\approx 5\times 10^5$ as the start time to count change in the momentum spread.
That is, we now examine
$\left[\expct{p^2}(t) -\expct{p^2}(t_2)\right]$.
Because at $t_2$, the population inside the regime $[-P_0,P_0]$ is about 0.88, we have
$P_c(t)=0.88\exp [-\Gamma (t-t_2)]$. Then we have
\begin{equation}
    \label{psquare2_CIII}
    \begin{split}
       \left[\expct{p^2}(t) -\expct{p^2}(t_2)\right] & \approx 0.88\ D_0 e^{-\Gamma (t-t_2)} (t-t_2) \\ 
			                                  & \quad + 0.12\ a (t-t_2)^2 \\
			                                  & \quad + a \int^t_{t_2} { R(t') (t-t')^2 \rmd t'},
		\end{split}
\end{equation}
where the first term account for the normal diffusion
as the trajectories diffuse from a chaotic sea to eventually land on a ballistic structure,
the second term
describes the ballistic transport for those trajectories
already outside the regime $[-P_0,P_0]$ at the start time $t_2$,
and the last term describes the impact on the transport dynamics due to
the population leakage from the regime $[-P_0,P_0]$, with
\begin{eqnarray}
R(t) &= & -\frac{d P_c}{dt} \nonumber \\
 &=& 0.88\ \Gamma \exp[-\Gamma (t-t_2] \approx  0.88\ \Gamma.
\end{eqnarray}
Equation~(\ref{psquare2_CIII}) thus suggests that once we reset the start time at $t_2$, there should be a normal diffusion stage, a transition stage due to the second term, followed by a superballistic transport period, i.e., $[\expct{p^2}(t) -\expct{p^2}(t_2)] \sim (t-t_2)^3$.   In Fig.~\ref{fig:K1_4M10_truncated} we present a numerical
log-log plot of $\left[\expct{p^2}(t) -\expct{p^2}(t_2)\right]$ vs $(t-t_2)$, in very good agreement with our analysis.
As a final note, the classical superballistic transport shown in Fig.~\ref{fig:K1_4M10_truncated} with the diffusion exponent $\nu\approx 3.1$ lasts very long, but this behavior
 cannot last forever.  In the end, almost all
trajectories from the initial ensemble will end up on ballistic structures and then purely ballistic transport will take over.

Returning to an early study~\cite{Matrasulov2005} of the classical
relativistic KR under the resonance condition, we have to disagree with 
some of their statements; from our results we conclude that 
classical superballistic
transport was not observed there because the investigation time scale
there was too short.


\section{Conclusions}

In this work, we 
show 
that both quantum and classical superballistic transport can occur in
a 
simple periodically driven system, namely, a relativistic
kicked-rotor system with a nonzero mass term.  To our knowledge, 
this appealing scenario has not been discussed before. 
Compared with previous
lattice-junction models for quantum superballistic transport, the
superballistic transport in our model occurs in 
momentum space as a consequence of a natural divide imposed by the relativistic dispersion: 
regions with low momentum 
effectively have a quadratic (bare)
dispersion relation (hence effectively a quasi-random on-site potential) and regions of high momentum 
effectively
have a linear (bare) dispersion relation (hence effectively a quasiperiodic potential).

Remarkably, though found in the same dynamical system, the quantum
superballistic and classical superballistic transport we have analyzed
are observed in much different parameter regimes. Indeed, in the
quantum case, the mechanism lies in the leakage of the quantum state
from a regime of dynamical localization to a regime of ballistic
transport.  This leakage can occur even when the underlying classical
limit has global KAM invariant curves separating the two regimes.  In
the classical case, it is necessary to break the global KAM curves
first to allow for leakage from a chaotic sea to ballistic
trajectories.
As a side result, we find that this kind of leakage in the classical
dynamics is unexpectedly complicated and further studies can be
motivated. For example, strictly speaking, the random patterns shown
in the panel (b) of Fig.~\ref{fig:PhaseSinglePoint} do not represent
chaos (chaos is defined as a positive Lyapunov exponent in the
asymptotic long-time limit, but this trajectory will eventually become
ballistic and hence has a zero Lyapunov exponent).  The detailed
characteristics of this type of irregular trajectory eventually
becoming a regular ballistic one deserve more attention.   The issue
of quantum-classical correspondence concerning this type of
trajectories is also of considerable interest for future studies.

The classical relativistic kicked rotor model may be realized by
considering relativistic electrons moving in the field generated by a
special electrostatic wavepacket~\cite{Chernikov1989}.  On the quantum
side,
a spinless version of the relativistic kicked rotor may be also
realized by considering a kicked
 tight-binding lattice whose on-site potential can be determined by
 the relativistic dispersion relation ~\cite{Gong2007,
   Monteiro}. However, due to the large time scales involved,  a
 direct observation of our numerical results reported here is
 unlikely.   As such it should be interesting enough to explore the
 system more to identify other signatures of superballistic transport
 at shorter time scales.




\begin{thebibliography}{99}%
\bibitem{review1} B. D. Hughes, Random Walks and Random Environments
(Clarendon Press, Oxford, 1995).
\bibitem{PRLbarkai}Y. He, S. Burov, R. Metzler, and E. Barkai, Phys. Rev.
Lett. {\bf 101}, 058101 (2008).
\bibitem{review2} R. Metzler and J. Klafter, Phys. Rep. {\bf 339}, 1 (2000).
\bibitem{review3} J. P. Bouchaud and A. Georges, Phys. Rep. {\bf 195}, 127 (1990).
\bibitem{review4} B. I. Henry, T.A.M. Langlands, and P. Straka, in Complex
Physical, Biophysical and Econophysical Systems, World
Scientific Lecture Notes in Complex Systems, edited by
R. L. Dewar and F. Detering (World Scientific, Singapore,
2010), Vol. 9.
\bibitem{exponential}J.~Wang, I. Guarneri, G. Casati, and J.~B. Gong, \prl\ {\bf 107}, 234104 (2011);
H.~L.~Wang, J.~Wang, I.~Guarneri, G.~Casati, and J.~B.~Gong, \pre\ {\bf 88}, 052919 (2013).
 \bibitem{abe}S. Abe and H. Hiramoto, Phys. Rev. A {\bf 36}, 5349 (1987); H.
Hiramoto and S. Abe, J. Phys. Soc. Jpn. {\bf 57}, 230 (1988);
{\bf 57}, 1365 (1988).
\bibitem{piechon} F. Pi\'{e}chon, Phys. Rev. Lett. {\bf 76}, 4372 (1996).
\bibitem{ketzmerick}R. Ketzmerick, K. Kruse, S. Kraut, and T. Geisel, Phys.
Rev. Lett. {\bf 79}, 1959 (1997).
\bibitem{Bao2007} K. L\"{u} and J.-D. Bao, Phys. Rev. E {\bf 76}, 061119 (2007).
\bibitem{Siegle2010} P. Siegle, I. Goychuk, P. Talkner, and P. H\"{a}nggi, Phys. Rev.
E {\bf 81}, 011136 (2010); P. Siegle, I. Goychuk, and P. H\"{a}nggi,
Phys. Rev. Lett. {\bf 105}, 100602 (2010); P. Siegle, I. Goychuk,
and P. H\"{a}nggi, Europhys. Lett. {\bf 93}, 20002 (2011).
\bibitem{Levi2012} L. Levi, Y. Krivolapov, S. Fishman and M. Segev, Nat. Phys. {\bf 8}, 912 (2012).
\bibitem{Hufnagel2001}L. Hufnagel, R. Ketzmerick, T. Kottos, and T. Geisel,
Phys. Rev. E 64, 012301 (2001).
\bibitem{Zhenjun2012}Z. J. Zhang, P. Q. Tong, J. B. Gong, and B. W. Li, \prl{\bf 108}, 070603 (2012).
\bibitem{Stutzer2013} S. St\"{u}tzer, T. Kottos, A. T\"{u}nnermann, S. Nolte,
D. N. Christodoulides, and A. Szameit,
Opt. Lett. {\bf 38}, 4675 (2013).
\bibitem{Chernikov1989} A. A. Chernikov, T. Tեl, G. Vattay, and G. M. Zaslavsky,
Phys. Rev. A {\bf 40}, 4072 (1989).
\bibitem{Matrasulov2005} D. U. Matrasulov, G. M. Milibaeva, U. R. Salomov, and
B. Sundaram, Phys. Rev. E {\bf 72}, 016213 (2005).
\bibitem{CasatiBook1995}G. Casati and B. V. Chirikov, Quantum Chaos: Between
Order and Disorder (Cambridge University Press, New
York, 1995).
\bibitem{Grempel1982}D. R. Grempel, S. Fishman, and R. E. Prange,
Phys. Rev. Lett. {\bf 49}, 833 (1982).
\bibitem{Prange1984}R. E. Prange, D. R. Grempel, and S. Fishman,
Phys. Rev. B {\bf 29}, 6500 (1982).
\bibitem{Berry1984} M. V. Berry, Physica D {\bf 10}, 369 (1984).
\bibitem{Simon1985} B. Simon, Annals of Physics {\bf 159}, 157 (1985).
\bibitem{Nomura1992} Y. Nomura, Y. H. Ichikawa, and W. Horton, Phys. Rev. A {\bf 45}, 1103 (1992).
\bibitem{Fishman}S. Fishman, D. R. Grempel, and R. E. Prange, Phys. Rev. Lett.
{\bf 49}, 509 (1982).
\bibitem{Mello1988} P. Mello, P. Pereyra, and N. Kumar,
Ann.~Phys. {\bf 181}, 290 (1988).
\bibitem{Evangelou1987} S. N. Evangelou and T. Ziman,
Journal of Physics C: Solid State Physics {\bf 20}, L235 (1987).
\bibitem{Sheng1996} D. N. Sheng and Z. Y. Weng,
Phys. Rev. B {\bf 54}, R11070 (1996).
\bibitem{Gong2007}J.~B.~Gong and J. Wang, Phys. Rev. E {\bf 76}, 036217 (2007).
\bibitem{Monteiro} T. Boness, S. Bose, and T. S. Monteiro, Phys. Rev. Lett. {\bf 96} ,
187201 (2006).

\end{thebibliography}
\end{document}